\documentclass[11pt]{article}
\usepackage{amssymb,amscd,amsmath}

\newcommand{\2}{{\hat sl}_2}

\newcommand{\pfc}{\Phi^{j}}
\newcommand{\bx}{\bar x}
\newcommand{\bz}{\bar z}

\newcommand{\gm}{\Gamma}

\newcommand{\ap}{\alpha_{+}}
\newcommand{\am}{\alpha_{-}}
\newcommand{\ip}{i^{\prime}}

\newcommand{\po}{\Phi^{j\,\bar j}}
\newcommand{\prh}{\rho^{\prime}}
\newcommand{\slc}{\Phi^j_{\mu.\bar \mu}}
\newcommand{\psf}{\Phi^{h.\bar h}_{q.\bar q}}
\newcommand{\psfa}{\Phi^h_q}
\newcommand{\ve}{\varepsilon}

\def\fzx#1{\Phi^{j_{#1}}(x_{#1},\bx_{#1},z_{#1},\bz_{#1})}
\def\dj#1{\Delta_{j_{#1}}}
\def\ff#1{\Phi^{j_{#1}}}

\def\h{\frac{1}{2}}
\def\j#1#2{j_{#1.#2}}
\def\C{\mathbb{C}}

\def\N{\mathbb{N}}

\def\j#1#2{j_{#1.#2}}

\def\cpf#1#2{\Phi_{#1.#2}}
\def\cpfp#1#2{\Phi_{#1.#2}^+}
\def\h{\frac{1}{2}}
\def\sn#1#2{\Phi^j_{#1.\bar #2}}
\def\mathid {{\mathchoice {{\rm 1\mskip-4mu l}} {{\rm 1\mskip-4mu l}}
{{\rm 1\mskip-4.5mu l}} {{\rm 1\mskip-5mu l}}}}
\def\sp{\boldsymbol{\phi}}
\def\boz{\boldsymbol z}
\def\boa{\boldsymbol \alpha}
\def\NP#1#2{ Nucl.Phys. B#1 (#2)}
\def\PL#1#2{ Phys.Lett. B#1 (#2)}
\def\CMP#1#2{ Commun.Math.Phys. #1 (#2)}

\title{ Some chiral rings of N=2 discrete superconformal series 
induced by SL(2) degenerate conformal field theories}

\author{Oleg Andreev\thanks{e-mail address: andreev@peterpan.ens.fr}
\thanks{On leave from Landau Institute for Theoretical Physics,
Moscow}\\ \\
Laboratoire de Physique Th\'eorique de l'\'Ecole Normale Sup\'erieure
\thanks{Unit\'e Propre du Centre National de la Recherche Scientifique,
associ\'ee \`a l'\'Ecole Normale Sup\'erieure et \`a l'Universit\'e 
de Paris-Sud.} ,\\
24 rue Lhomond, 75231 Paris C\'EDEX 05, France}

\date{}
\begin{document}
\maketitle

\vspace{-10cm}
\begin{flushright}
hep-th/9612043

{\sf LPTENS-96/68}
\end{flushright}

\vspace{8cm}
\begin{abstract}
By generalizing a fermionic construction, a natural relation is found
between $SL(2)$ degenerate conformal field theories and some $N=2$
discrete superconformal series. These non-unitary models contain, as a 
subclass, $N=2$ minimal models. The construction permits one to
investigate the properties of chiral operators in the $N=2$ models. A
chiral ring reveals a close connection with underlying quantum group 
structures.
\end{abstract}

\section{Introduction}

Recent discussion of M-theory and string dualities involve $N=2$ two
dimensional superconformal field theories \cite{MKL}. Superstrings
with $N=2$ algebra on the world sheet were shown to describe self-dual
Yang-Mills and gravity in a K\"ahler space-time with signature (2.2) (
see e.g. \cite{Ma} and references therein ). Such strings are the
exactly solvable four dimensional string theories. Models of string
compactification based on $N=2$ superconformal models are also known
from the work 
by Gepner \cite{G}. Their key stones are the so-called $N=2$ 
minimal models \cite{BF}. The latter are a subclass of $N=2$ discrete
series \cite{DP}. These models are non-unitary and have, in general, an
OP algebra of
primary fields which is not closed. Nevertheless the presence of
singular vectors in the representations of $N=2$ algebra for such
series provides a strong evidence for exact solvability. One
motivation for the present work was to do a step towards an exact
solution using the recent progress with $SL(2)$ degenerate
conformal field theories \cite{A1}.

Another motivation was to try to understand the nature of chiral rings
\cite{L}. In fact, it is the simplest structure of $N=2$
superconformal theories. At first sight, it is rather difficult to
extract an origin of chiral rings because the same quantum numbers are
shared by conformal dimensions, U(1) charges and weights of quantum
group. So I am bound to learn something if I succeed.

The outline of the paper is as follows.

In sections 2.1 and 2.2, a brief overview of the essential elements
of $SL(2)$ and $N=2$ two dimensional conformal field theories is
given. I will mainly 
concentrate on a degenerate case when representations
contain the so-called singular vectors. In section 2.3 I describe a
relation between some of these models, and the use of the fermionic
construction of Di Vecchia, Petersen, Yu and Zheng to obtain a proper
relation between correlation functions. In particular, I show that the
construction works for any complex level $k$ of $\2$. The main body of
this work is presented in sections 3.1 and 3.2. In section 3.1 by
computing the OP algebra of primary chiral fields I show that they
don't generate a ring structure. The origin of this disaster is the
non-unitarity of the models. In the case at hand the U(1) conservation
law doesn't provide a proper selection rule. It forces one to look for
more fine structures. The solution of the problem is given in section
3.2 by introducing Moore-Reshetikhin operators \cite{MR}. This
provides a strong evidence for a quantum group nature of chiral rings. 
Section 4 will present my conclusions and some open problems. In the 
appendices I give technical details which are relevant for the explicit
construction of the chiral rings.

\section{Preliminaries}

{\bf 2.1 SL(2) degenerate conformal field theories}
\renewcommand{\theequation}{2.\arabic{equation}}
\setcounter{equation}{0}

\vspace{.5cm}
The theories have $\2\times\2$ algebra as the symmetry algebra. The
holomorphic form of currents has the following OP algebra \cite{FZ}:
\begin{align}
J(x_1,z_1)J(x_2,z_2)=-k\frac{x_{12}^2}{z_{12}^2}-2\frac{x_{12}}{z_{12}}
J(x_2,z_2)-\frac{x_{12}^2}{z_{12}}\frac{\partial}{\partial x_2}J(x_2,z_2)+O(1)
\qquad,
\end{align}
where $k$ is the level of the algebra, $z$ - a point on the
sphere, $x$ is an
isotopic coordinate of $sl_2$
\footnote{The generators of $sl_2$ look like 
$S_j^-=\frac{\partial}{\partial x}\,,\,
S_j^0=-x\frac{\partial}{\partial x}+j\,,\,
S_j^+=-x^2\frac{\partial}{\partial x}+2jx$.},
\newline $z_{ij}=z_i-z_j,\,x_{ij}=x_i-x_j$. The
same OP expansion, of course, is valid for the antiholomorphic form
\footnote{I will not write down antiholomorphic OP expansions when
their form follows from holomorphic one.}.

The standard holomorphic currents are the Taylor series components
of $J(x,z)$
\begin{align}
J(x,z)=J^+(z)-2xJ^0(z)-x^2J^-(z)\qquad.
\end{align}
They form the OP algebra:
\begin{align}
J^{\alpha}(z_1)J^{\beta}(z_2)=\frac{k}{2}\frac{q^{\alpha\beta}}{z_{12}^2}
+\frac{f^{\alpha\beta}_{\gamma}}{z_{12}}
J^{\gamma}(z_2)+O(1)\qquad,
\end{align}
where $q^{00}=1,\,\,q^{+-}=q^{-+}=2,\,\,
f^{0+}_{+}=f^{-0}_{-}=1,\,\,f^{+-}_{0}=2;\,\,\alpha,\beta=0,+,-$. It
is a little exercise in OP expansions to derive (2.3) from (2.1) and
vice versa.

The stress-energy tensor of the model has two independent components which can
be chosen in the Sugawara form
\begin{align}
\begin{split}
T(z)=\frac{1}{k+2}q_{\alpha\beta}:J^{\alpha}(z)J^{\beta}(z):\qquad,\\
\bar T(\bz)=\frac{1}{k+2}q_{\alpha\beta}:\bar
J^{\alpha}(\bz)\bar J^{\beta}(\bz):\qquad.
\end{split}
\end{align}
It is known that each component provides the Virasoro algebra with the
central charge $c=\frac{3k}{k+2}$.

Define the primary fields as
\begin{align}
J(x_1,z_1)\po(x_2,\bx_2,z_2,\bz_2)=-2j\frac{x_{12}}{z_{12}}
\po(x_2,\bx_2,z_2,\bz_2)-\frac{x_{12}^2}{z_{12}}\frac{\partial}{\partial x_2}
\po(x_2,\bx_2,z_2,\bz_2)+O(1)\quad.
\end{align}
\newline It should be noted that in the general case the primary fields are
non-polynomial in $x,\bx$. Furthermore, $J(x,z),\,\bar J(\bx,\bz)$ are
not primary.

The complete system of fields involved in the theory includes, besides
the primary fields $\po$, all the fields (descendants) of the form
\begin{align}
J^{\alpha_1}_{n_1}(x)\dots J^{\alpha_N}_{n_N}(x)
\bar J^{\beta_1}_{\bar n_1}(\bx)\dots\bar J^{\beta_M}_{\bar n_M}(\bx)
\po(x,\bx,z,\bz)\qquad,
\end{align}
where $J^{\alpha}_{n}(x),\,\,\bar J^{\beta}_{\bar n}(\bx)$ are the
Laurent series
components of $J(x,z)\text{ and }\bar J(\bx,\bz)$, respectively. From a
mathematical point of view the primary fields correspond to the highest weight 
vectors of $\2\times\2$. As to the parameters $j$'s, they are
the weights of the representations.

I will consider only the diagonal embedding the physical space of
states into a tensor product of holomorphic and antiholomorphic 
sectors. Such models are
known as "A" series. Since for these models all primary fields are spinless,
i.e. $\bar j\equiv j(\bar \Delta\equiv\Delta)$, I suppress $\bar j$-dependence
below.

In \cite{KK} Kac and Kazhdan found that the highest weight representation
of $\2$ is reducible if the highest weight $j$ takes the values $\j{n}{m}$
defined by
\begin{align}
j^+_{n.m}=\frac{1-n}{2}(k+2)+\frac{m-1}{2}\qquad\text{ or }\qquad
j^-_{n.m}=\frac{n}{2}(k+2)-\frac{m+1}{2}\quad,
\end{align}
with $k\in\C\,,\,\{n,m\}\in\N$. Note that the unitary representations 
are a subset of
the Kac-Kazhdan set namely, they are given by $j^+_{1.m}$ with the
integer level $k$.

I will call SL(2) conformal field theories with the primary fields
parametrized by the Kac-Kazhdan list as the degenerate SL(2) conformal
field theories.  

The Operator Product of any two operators is given by
\begin{align}
\phi^{j_1}(x,\bx,z,\bz)\phi^{j_2}(0,0,0,0)=\sum_{j_3}C^{j_1\,j_2}_{j_3}
(x,\bx,z,\bz)\phi^{j_3}(0,0,0,0)\quad.
\end{align}

It is well-known that all the coefficient functions $C^{j_1\,j_2}_{j_3}
(x,\bx,z,\bz)$ in the expansion (2.8) can be expressed via the 
weights (conformal dimensions) of the primary fields (basic operators)
and the structure constants
of Operator Algebra \cite{BPZ}. The structure constants are defined as
coefficients at the primary fields in the OP expansion
\begin{align}
\ff{1}(x,\bx,z,\bz)\ff{2}(0,0,0,0)=\sum_{j_3}\frac{|x|^{2(j_1+j_2-j_3)}}
{|z|^{2(\dj{1}+\dj{2}-\dj{3})}}C^{j_1\,j_2}_{j_3}\ff{3}(0,0,0,0)\qquad.
\end{align}

The normalized two and three point functions of the primary fields can be
represented as
\begin{align}
\begin{split}
\langle\fzx{1}\fzx{2}\rangle=&
\delta^{j_1j_2}\frac{|x_{12}|^{4j_1}}{|z_{12}|^{4\dj{1}}}\qquad,\\
\langle\fzx{1}\fzx{2}\fzx{3}\rangle=&C^{j_1\,j_2\,j_3}
\prod_{n<m}\frac{|x_{nm}|^{2\gamma_{nm}(j)}}{|z_{nm}|^{2\gamma_{nm}(\Delta)}}\quad,
\end{split}
\end{align}
where $\gamma_{12}(y)=y_1+y_2-y_3\,,\,
\gamma_{13}(y)=y_1+y_3-y_2\,,\, \gamma_{23}(y)=y_2+y_3-y_1\,\,\,
\text{and}\,\,\Delta_j=\frac{j(j+1)}{k+2}$.

As to the four point function, one can find it in the following form \cite{FZ}
\begin{align}
\begin{split}
\langle\fzx{1}\dots\fzx{4}\rangle
=G^{j_1,j_2,j_3,j_4}(x,\bx,z,\bz)
\prod_{n<m}\frac{|x_{nm}|^{2\ve_{nm}(j)}}{|z_{nm}|^{2\ve_{nm}
(\Delta)}}\quad,
\end{split}
\end{align}
with
$\ve_{14}(y)=2y_1,\,\ve_{23}(y)=y_1+y_2+y_3-y_4,\,
\ve_{24}(y)=-y_1+y_2-y_3+y_4,\,$
\newline$\ve_{34}(y)=-y_1-y_2+y_3+y_4$  and
\begin{align*}
x=\frac{x_{12}x_{34}}{x_{14}x_{32}}\,,\qquad\bx=\frac{\bx_{12}\bx_{34}}
{\bx_{14}\bx_{32}}\,,\qquad z=\frac{z_{12}z_{34}}{z_{14}z_{32}}\,,\qquad
\bz=\frac{\bz_{12}\bz_{34}}{\bz_{14}\bz_{32}}\qquad.
\end{align*}

The functions
$G^{j_1,j_2,j_3,j_4}(x,\bx,z,\bz)$ are given by (see \cite{A1} for details)
\begin{align}
\begin{split}
G^{j_1,j_2,j_3,j_4}(x,\bx,z,\bz)&=Z(j_1,j_2,j_3,j_4)|z|^a|1-z|^b
\prod_{i=1}^{n_1-1}\prod_{\ip=1}^{m_1-1}\int d^2\,u_i\int d^2\,w_{\ip}\,\,
\vert u_i-w_{\ip}\vert^{-4}\times\\
&\times\prod_{i=1}^{n_1-1}\vert u_i\vert^{4\alpha_1\am}\vert
1-u_i\vert^{4\alpha_2\am} \vert x-u_i\vert^{4\alpha_{21}\am}\vert
z-u_i\vert^{4\alpha_3\am} \prod_{i<\ip}^{n_1-1}\vert
u_{i\ip}\vert^{4\am^2} \times\\
&\times\prod_{i=1}^{m_1-1}\vert
w_{i}\vert^{4\alpha_1\ap}\vert 1-w_i\vert^{4\alpha_2\ap}\vert
x-w_{i}\vert^{4\alpha_{21}\ap} \vert z-w_i\vert^{4\alpha_3\ap}
\prod_{i<\ip}^{m_1-1}\vert w_{i\ip}\vert^{4\ap^2}.
\end{split}
\end{align}
Here $a=4j_1j_2\ap^2\,,\,b=4j_1j_3\ap^2\,,\,\am=-\sqrt{k+2}\,,\,\ap\am=-1\,,\,
\alpha_i$'s are defined via $\alpha_i=\frac{1-N_i}{2}\am+
\frac{1-M_i}{2}\ap$. It should be noted that $N_i$'s ($M_i$'s)
are linear combinations of $n_i$'s ($m_i$'s)
and their form depends on the parametrizations (2.7).

In order to take into account a relative normalization between the 
operators of
the Dotsenko-Fateev models and the ones of the SL(2) degenerate conformal
field theories one has to introduce the normalization constants
$Z(j_1,j_2,j_3,j_4)$. For their explicit form I refer to the
original work \cite{A1}.

{}From the set (2.7) it is worth distinguishing the so-called admissible
representations \cite{KW}, which correspond to the rational level k. In the
case $k=-2+p/q$, with the coprime integers $p$ and $q$, it is possible to
recover the minimal models (series with $c<1$, \cite{BPZ}) via the 
Drinfeld-Sokolov reduction. On the other hand $k=-2-p/q$ leads to the 
Liouville series with
$c>25$. The second point is finite dimensional representations of the
modular group for such representations.

At the rational level $k=-2+p/q$ there is a symmetry
$j^-_{n,m}=j^+_{q-n+1,p-m}$ which allows one to reduce the fields
parameterized by $j^-_{n.m}$ to the fields parameterized by $j^+_{n.m}$.
In this case the structure constants of the Operator Product algebra 
are given by
\begin{align}
\begin{split}
&C^+(n_1,m_1;n_2,m_2;n_3,m_3) =\\
&\frac{\Gamma^{\h}[\rho]}{\Gamma^{\h}[1-\rho]}
P(\sigma^{\prime}-\h,\sigma+\h)
 \prod_{\{1,2,3\}}(-)^{\frac{n_i-1}{2}}\rho^{(1-n_i)}
\biggl(\frac{\Gamma[n_i-m_i\rho]}{\Gamma[1-n_i+m_i\rho]}\biggr)^{\h}
\frac{P(\sigma^{\prime}-n_i+\h,\sigma-m_i+\h)}{P(n_i,m_i)}\,,
\end{split}
\end{align}
\begin{align}
C^-(n_1,m_1;n_2,m_2;n_3,m_3)=
\rho^{-\h}P(\sigma^{\prime},\sigma+\h)\prod_{\{1,2,3\}}
\rho^{-(n_i-1)(m_i-\h)}P(\sigma^{\prime}-n_i,\sigma-m_i+\h)\quad.
\end{align}
Here $\sigma^{\prime}=\frac{n_1}{2}+\frac{n_2}{2}+\frac{n_3}{2}\,,\,
\sigma=\frac{m_1}{2}+\frac{m_2}{2}+\frac{m_3}{2}\,,\,\rho=\alpha_+^2\,\,
\text{and}\,\,\prh=\alpha_-^2\,$. The function 
$P(n,m)$ is defined by
\begin{gather*}
P(n,m)=\prod_{i=1}^{n-1}\prod_{j=1}^{m-1}[i\prh-j]^{-2}
\prod_{i=1}^{n-1}\frac{\Gamma[i\prh]}{\Gamma[1-i\prh]}
\prod_{j=1}^{m-1}\frac{\Gamma[j\rho]}{\Gamma[1-j\rho]}\quad ,\quad
P(1,1)=1\quad.
\end{gather*}
It should be noted that $n_3\,,\,m_3$ in (2.14) belong to the field
parameterized by $j^-_{n.m}$. Such choice clarifies the quantum group 
structure $(U_qosp(2/1),U_qsl(2))$  of the model \cite{MF}.

It is easy to see from (2.13) and (2.14) that the OP algebra at the rational
level is closed in the grid $1\leq n_i\leq q,\,\,1\leq m_i\leq p-1$.  The
corresponding fusion rules are given by 
\begin{align} 
\begin{cases}
\vert n_{12}\vert +1\,\leq\, n_3\leq\text{min}\,(n_1+n_2-1,\,
2q-n_1-n_2+1)\,,\,\,\text{with}\,\,\,\Delta n_3=1\quad,\\
\vert m_{12}\vert +1\,\leq\, m_3\leq\text{min}\,(m_1+m_2-1,\,
2p-m_1-m_2-1)\,,\,\,\text{with}\,\,\,\Delta m_3=2\quad.
\end{cases}
\end{align}
In the above $\Delta$ means a step. These fusion rules 
were first found in \cite{FZ,AY} from the differential equations for the 
conformal blocks.

Let me now define the primary fields of the algebra (2.3) via 
$\ff{}(x,\bx,z,\bz)$ as
\begin{align}
\slc(z,\bz)=\frac{1}{\mathcal N(j,\mu,\bar \mu)}
\oint_C \oint_{\bar C}dx\,d\bx\,\, x^{\mu-1-j}\,\,\bx^{\bar
\mu-1-j}\ff{}(x,\bx,z,\bz)\quad,
\end{align}
where $C,\bar C$ are closed contours, $\mu,\bar \mu$ are arbitrary 
parameters. The normalization factors $\mathcal N(j,\mu,\bar \mu)$ are
computed in Appendix A. Explicitly
\begin{align}
\mathcal N(j,\mu,\bar \mu)=\gm[2j+1]\{\gm[1+j+\mu]\gm[1+j-\mu]
\gm[1+j+\bar \mu]\gm[1+j-\bar \mu]\}^{-\frac{1}{2}} \quad.
\end{align}

Using the OP expansion (2.5) as well as (2.2) one arrives at
\begin{equation}
\begin{split}
J^0(z_1)\slc(z_2,\bz_2)=&\frac{\mu}{z_{12}}\slc(z_2,\bz_2)+O(1)\quad,\\
J^{\pm}(z_1)\slc(z_2,\bz_2)=&
\frac{1}{z_{12}}\mathcal M_{\pm}
\Phi^j_{\mu\pm 1.\bar \mu}(z_2,\bz_2)+O(1)\quad,
\end{split}
\end{equation}
with $\mathcal M_{\pm}=(j\pm\mu+1)\mathcal N(j,\mu\pm 1,\bar \mu)/
\mathcal N(j,\mu,\bar \mu)$.

The highest(lowest) weight vectors of $\2\times\2$ algebra can be
extracted from (2.16) by setting $\mu=\bar \mu=j\,(\mu=\bar
\mu=-j)$. This is an immediate consequence of (2.17) and (2.18).

Before discussing the $N=2$ discrete superconformal field theories, I
pause here to emphasize one important point. The primary fields
defined in (2.16) depend on contours $C_i(\bar C_i)$ in the isotopic
spaces. From this point of view one has the non-local
operators. However it hasn't influence on the main results obtained below.

\vspace{.3cm}
{\bf 2.2 N=2 discrete superconformal field theories}
\vspace{.3cm}

The theory has $N=2\times N=2$ algebra as the symmetry algebra.
The holomorphic part, $N=2$ superconformal algebra, is generated by four
local currents: $T(z)\,,\,G^{}(z)\,\,\text{and}\,\,J(z)$. The
fermionic currents $G^{}(z)$ have a conformal dimension
$(\frac{3}{2},0)$, and the bosonic currents
$J(z)\,\,\text{and}\,\,T(z)$ have $(1,0)$ and $(2,0)$,
respectively. The current $T(z)$ is the holomorphic stress-energy tensor.

The algebra is determined by the following Operator Product
expansions:
\begin{align}
\begin{split}
&J(z_1)J(z_2)=\frac{c_2/4}{z_{12}^2}+O(1)\quad,\quad
T(z_1)J(z_2)=\frac{1}{z_{12}^2}J(z_2)+\frac{1}{z_{12}}
\frac{\partial}{\partial z_2}J(z_2)+O(1)\quad,\\
&J(z_1)G^{\pm}(z_2)={\pm}\frac{1/2}{z_{12}}G^{\pm}(z_2)+O(1)\quad,\quad
T(z_1)G^{\pm}(z_2)=\frac{3/2}{z_{12}^2}G^{\pm}(z_2)+
\frac{1}{z_{12}}\frac{\partial}{\partial z_2}G^{\pm}(z_2)+O(1)\,\,,\\
&T(z_1)T(z_2)=\frac{3c_2/2}{z^4_{12}}+\frac{2}{z^2_{12}}T(z_2)+
\frac{1}{z_{12}}\frac{\partial}{\partial z_2}T(z_2)+O(1)\,\,,
{}\\
&G^+(z_1)G^-(z_2)=\frac{2c_2}{z_{12}^3}+\frac{4}{z_{12}^2}J(z_2)+
\frac{2}{z_{12}}(\frac{\partial}{\partial z_2}J(z_2)+T(z_2))+O(1)\quad,
{}
\end{split}
\end{align}
The central charge $c_2$ is related to the usual Virasoro($N=0$)
central charge $c$ by $c_2=c/3$. The normalization is fixed so that
$c_2=1$ for the free scalar superfield.

The three sectors of the theory are given by three moddings of the
generators, corresponding to three ways of choosing boundary conditions
on the cylinder. Because I am interested in chiral rings let me
restrict to the Neveu-Schwarz (NS) sector\footnote{Notice that it is
possible to recover ground states of the Ramond sector from the NS
sector under the spectral flow mapping \cite{S}.}. This sector has integer
modes for the bosonic currents, but half-integers for fermionic ones.

The corresponding primary fields are given by
\begin{equation}
\begin{split}
&J(z_1)\psf(z_2,\bz_2)=\frac{q}{z_{12}}\psf(z_2,\bz_2)+O(1)\quad,\\
&T(z_1)\psf(z_2,\bz_2)=\frac{h}{z_{12}^2}\psf(z_2,\bz_2)
+\frac{1}{z_{12}}\frac{\partial}{\partial z_2}\psf(z_2,\bz_2)+O(1)\quad,\\
&G^{\pm}(z_1)\psf(z_2,\bz_2)=\frac{1}{z_{12}}G_{-\frac{1}{2}}^{\pm}
\psf(z_2,\bz_2)+O(1)\quad.
\end{split}
\end{equation}
Here $h$ and $q$ are a conformal dimension and U(1) charge.

The complete system of fields involved in the theory is obtained by
acting with the all negative frequency modes of the currents on the
primary fields. From mathematical point of view the primary fields 
correspond to the
highest weight vectors of $N=2\times N=2$ algebra.

As in the previous section I will only consider the diagonal
embeddings of the physical space of
states into a tensor product of holomorphic and antiholomorphic
sectors, the so-called "A" series, and due to this reason I will suppress 
$\bar h,\bar q$-dependence below.

It is known \cite{BF,DP} that the highest weight representation is
reducible if the conformal dimension takes the values defined by
\begin{align}
&h^I_{n.m}=\frac{1}{4(k+2)}\bigl[\bigl(m-(k+2)(n-1)\bigr)^2-1\bigr]
-(k+2)q^2 \quad,\quad n,m\in\N\quad,\\
&h^{II}_p=\frac{1}{4(k+2)}(p^2-1)\,\pm\,pq\quad,\quad p\in\N\quad,
\end{align}
where $c_2=1-\frac{2}{k+2}\,,\,\,k\in\C$.

I will call theories with the primary fields parametrized by this set
as the discrete $N=2$ superconformal models. Note that the unitary
minimal models are a subset of the discrete ones namely, they are
given by $h^I_{1.m}$ with the integer parameter $k$.

\vspace{.3cm}
{\bf 2.3 N=2 via SL(2) degenerate conformal field theories}

\vspace{.3cm}
In order to write down correlation functions of the $N=2$ discrete 
superconformal field theories it seems very natural to use 
the fermionic construction proposed by Di Vecchia,
Petersen, Yu and Zheng to build the unitary representations of
the $N=2$ superconformal algebra in terms of free fermions and unitary
representations of $\2$ \cite{DP}. In fact one can do better:
the only difference between the unitary representations of $\2$ and
degenerate ones is a value of $k$ (see (2.7)). Therefore one can relate
the degenerate representations of $\2$ to the discrete representations
of $N=2$. 

Let me sketch the main points of this construction. The
holomorphic part is described in terms of the free fermions
$\psi^{\pm}(z)$ and $\2$ algebra. The U(1) current and
stress-energy tensor of the fermions are given by
\begin{gather*}
j(z)=:\psi^+(z)\psi^-(z):\qquad,\qquad 
T_{\psi}=\h :j(z)j(z):\qquad.
\end{gather*}

It is straightforward to see that in the case of a general $k$ the
$N=2$ currents are also expressed as \cite{DP}
\begin{equation}
\begin{split}
&J(z)=\frac{1}{2(k+2)}\bigl(2J^0(z)+kj(z)\bigr)\quad,\quad
G^{\pm}(z)=\sqrt{\frac{2}{k+2}}\psi^{\pm}(z)J^{\pm}(z)\quad, \\
&T(z)=T_{\2}+T_{\psi}-\frac{1}{k+2}:\bigl(J^0(z)-j(z)\bigr)^2:\quad,
\end{split}
\end{equation}
where $T_{\2}(z)$ is the Sugawara stress-energy tensor given by
(2.4). The OP expansions of $J^{\alpha}(z)$ are defined in (2.3).

The primary fields of the $N=2$ superconformal theories can be written as 
\begin{gather}
\psfa(z,\bz)=\Phi^j_{\mu}(z,\bz)\,\mathid\quad.
\end{gather}
Here $\mathid$ is a trivial field (identity operator) which
corresponds to the vacuum 
of the fermionic system in the NS sector and $\Phi^j_{\mu}$'s are the
primaries of $\2\times\2$. 

The conformal dimensions and U(1) charges are expressed via $j$ and $\mu$ as
\begin{gather}
h=\frac{j(j+1)}{k+2}-\frac{\mu^2}{k+2}\quad
,\quad\quad\quad q=\frac{\mu}{k+2}\quad.
\end{gather}

To give a relation between correlation functions of the above models,
let me now proceed in complete accordance with the derivation of the
Knizhnik-Zamolodchikov (KZ) equations \cite{KZ}. Inserting the 
constraint\footnote{Since the field $\mathid$ has trivial OP expansions
I omit modes of $T_{\psi}(z)$ and $j(z)$.} 
\begin{align*}
(k+2)L_{-1}=
\sum_{n=-\infty}^{+\infty}g_{\alpha\beta}:J_{n}^{\alpha}J_{-1-n}^{\beta}:
-\sum_{n=-\infty}^{+\infty}J_{n}^0J_{-1-n}^0 
\end{align*}
into a correlation function, I find
\begin{align}
(k+2)\frac{\partial}{\partial z_i}
\langle\,\prod_{i=1}^{N}\Phi^{h_i}_{q_i}(z_i,\bz_i)\,\rangle
=\sum_{i\neq j}^N g_{\alpha\beta}\frac{t^{\alpha}_it^{\beta}_j}{z_{ij}}
\langle\,\prod_{i=1}^{N}\Phi^{h_i}_{q_i}(z_i,\bz_i)\,\rangle
-2\sum_{i\neq j}^N\,\frac{\mu_i\mu_j}{z_{ij}}
\langle\,\prod_{i=1}^{N}\Phi^{h_i}_{q_i}(z_i,\bz_i)\,\rangle\quad.
\end{align}
Here $t^{\alpha}_i$'s are generators of $sl(2)$.
 
The solution of the above equations is given by
\begin{align}
\langle\,\prod_{i=1}^{N}\Phi^{h_i}_{q_i}(z_i,\bz_i)\,\rangle
=\prod_{i<j}\vert z_{ij}\vert^{\frac{-4\mu_i\mu_j}{k+2}}
\langle\,\prod_{i=1}^{N}\Phi^{j_i}_{\mu_i}(z_i,\bz_i)\,\rangle\quad,
\end{align}
where the last factor is a solution of the standard KZ equations for 
the $SL(2)$ conformal field theory, namely
\begin{align*}
(k+2)\frac{\partial}{\partial z_i}
\langle\,\prod_{i=1}^{N}\Phi^{j_i}_{\mu_i}(z_i,\bz_i)\,\rangle
=\sum_{i\neq j}^N g_{\alpha\beta}\frac{t^{\alpha}_it^{\beta}_j}{z_{ij}}
\langle\,\prod_{i=1}^{N}\Phi^{j_i}_{\mu_i}(z_i,\bz_i)\,\rangle\quad.
\end{align*}
 
So I obtain the relation between the correlation functions. Let me
conclude this section by giving a few remarks.
\newline (i) It is clear from (2.7) and (2.25) that one can 
recover only the first 
degenerate series $h^I$ of $N=2$ superconformal algebra via the degenerate
representations of $\2$\footnote{This is the case for correlation functions
too.}. However, the primary fields parametrized by the first series
$h^I$ form a closed OP algebra, i.e. there is decoupling of the second
series $h^{II}$. To see this, it is convenient to use the free field
representation. More discussion on this point is given in Appendix B.
\newline (ii) In the case of $N=2$ unitary minimal models it is
possible to derive the relation (2.27) via the Fateev-Zamolodchikov
parafermions \cite{FZ2,M}. However for a non-integer parameter $k$ the
algebra of the parafermionic currents is not closed and leads to an
ill-defined parafermionic theory. On the other hand, there is a strong
indication on a finite number of order parameters in such
''parafermionic theory '' for a rational $k$ because a proper $SL(2)$
theory has the closed OP algebra of the primary fields in this case.

\section{Chiral rings}

{\bf 3.1 Primary chiral fields}
\renewcommand{\theequation}{3.\arabic{equation}}
\setcounter{equation}{0}

\vspace{.5cm}
Among the primary fields of the Neveu-Schwarz sector of $N=2$ models 
it is worth to distinguish the so-called primary chiral fields
introduced by Lerche, Vafa and Warner in \cite{L}. 
Such fields satisfy, in addition to (2.20), the condition
\begin{gather}
G^+_{-\frac{1}{2}}\psfa(z,\bz)=0\quad.
\end{gather}
The anti-chiral fields are defined by replacing
$G^+_{-\frac{1}{2}}\rightarrow G^-_{-\frac{1}{2}}$.
 
Using (2.19) one can deduce that for such states $h=q$. The equations
(2.21-2.22) allow me to find the conformal dimensions in terms of integers as
\begin{alignat}{2}
&h_1^I=\frac{1-n}{2}+\frac{m-1}{2(k+2)}\quad, &\qquad\quad
&h_2^I=\frac{n-1}{2}-\frac{m+1}{2(k+2)}\quad;\\
&h^{II}_1=\frac{p-1}{4(k+2)}\qquad, &\qquad\quad
&h_2^{II}=-\frac{p+1}{4(k+2)}\quad.
\end{alignat}

On the other hand, the relationship between the $SL(2)$ and $N=2$
models implies that the
primary chiral fields correspond to the highest weight vectors of the 
$\2\times\2$ algebra. Note that a solution $\mu=-j-1$ of equations
(2.25) with $h=q$ is forbidden because it corresponds to a zero norm 
state (see (2.17)). As a result, one has the following set of the
conformal dimensions provided by $SL(2)$
\begin{align}
h_{n.m}^+=\frac{1-n}{2}+\frac{m-1}{2(k+2)}\quad, \quad\quad
h_{n.m}^-=\frac{n}{2}-\frac{m+1}{2(k+2)}\quad.
\end{align}
It is evident that for a general $k$ it is possible to recover
dimensions: $h_1^I\,,\,\,h_2^I\,\,\text{with}\,\,n>1$ and
$h_1^{II}$ with odd $p$. The other solutions are decoupled. The second
series decoupling is discussed in Appendix B. As to $h_2^I$ with
$n=1$, it is usual zero vectors decoupling in 2d conformal field
theories.

Since $h^{\pm}$ are parameterized by two integers $(n,m)$ it is useful
to denote the primary chiral fields $\Phi^{h^{\pm}}_{h^{\pm}}(z,\bz)$
as $\Phi^{\pm}_{n.m}(z,\bz)$.

The correlation functions of the primary chiral fields 
parameterized by (3.4) are computable by the relation (2.27) in terms 
of the correlation functions of the highest weight vectors. For
instance, a small calculation shows that the four point function of 
$\cpfp{n}{m}$ is given by
\begin{align}
\begin{split}
{}&\langle
\cpfp{n_1}{m_1}(z_1,\bz_1)\cpfp{n_2}{m_2}(z_2,\bz_2)
\cpfp{n_3}{m_3}(z_3,\bz_3)\Phi^{\dagger}_{n_4.m_4}(z_4,\bz_4)\rangle
=Z(h_1,h_2,h_3,h_4)\prod_{i=1}^3 \vert z_{i4}\vert^{-4h_i}\times\\
&\times
\prod_{i=1}^{3}\oint_{C_i}\frac{dx_i}{x_i}
\oint_{{\bar C}_i}\frac{d\bx_i}{\bx_i}
\oint_{C_4}\frac{dx_4}{x_4^{1+2\am^2h_4}}
\oint_{{\bar C}_4}\frac{d\bx_4}{\bx_4^{1+2\am^2h_4}}
\prod_{i<\ip}\vert x_{i\ip}\vert^{2\am^2\ve_{i\ip}(h)}
\prod_{i=1}^{n_1-1}\int d^2u_i\times \\
&\times
\prod_{\ip=1}^{m_1-1}\int d^2w_{\ip}
\prod_{i=1}^{n_1-1}\vert u_i\vert^{4\alpha_1\am}
\vert1-u_i\vert^{4\alpha_2\am} 
\vert x-u_i\vert^{4\alpha_{2.1}\am}
\vert z-u_i\vert^{4\alpha_3\am} \prod_{i<\ip}^{n_1-1}\vert
u_{i\ip}\vert^{4\am^2} \times\\
&\times
\prod_{i=1}^{m_1-1}
\vert w_{i}\vert^{4\alpha_1\ap}
\vert 1-w_i\vert^{4\alpha_2\ap}
\vert x-w_i\vert^{4\alpha_{2.1}\ap}
\vert z-w_i\vert^{4\alpha_3\ap}
\prod_{i<\ip}^{m_1-1}\vert w_{i\ip}\vert^{4\ap^2}
\prod_{i=1}^{n_1-1}\prod_{\ip=1}^{m_1-1}\vert u_i-w_{\ip}\vert^{-4}\quad.
\end{split}
\end{align}
Here $\alpha_i=\frac{1-N_i}{2}\am+\frac{1-M_i}{2}\ap$ with
\begin{alignat*}2
N_1 &=\frac{n_1}{2}+\frac{n_2}{2}-\frac{n_3}{2}-\frac{n_4}{2}\quad; &
\quad M_1
&=\frac{m_1}{2}+\frac{m_2}{2}-\frac{m_3}{2}-\frac{m_4}{2}\quad ; \\
N_2 &=\frac{n_1}{2}-\frac{n_2}{2}+\frac{n_3}{2}-\frac{n_4}{2}\quad; &
\quad M_2
&=\frac{m_1}{2}-\frac{m_2}{2}+\frac{m_3}{2}-\frac{m_4}{2}\quad ; \\
N_3 &=\frac{n_1}{2}+\frac{n_2}{2}+\frac{n_3}{2}+\frac{n_4}{2}-1\quad; &
\quad M_3
&=\frac{m_1}{2}+\frac{m_2}{2}+\frac{m_3}{2}+\frac{m_4}{2}\quad.
\end{alignat*}
A conjugate field $\Phi^{\dagger}_{n.m}$ is defined as 
$\Phi^{\dagger}_{n.m}(z,\bz)=\Phi^h_{-h}(z,\bz)$. Note that the U(1)
conservation law provides $h_4=h_1+h_2+h_3$.

One can try to analyze singularities of (3.5) in order to learn the OP
algebra of the primary chiral fields. However, due to the contours
$C_i(\bar C_i)$, it is a difficult task. On the other hand, it is enough
to set $n_1=m_1=1$ into a 4-point function 
\begin{align*}
\langle
\cpfp{n_1}{m_1}(z_1,\bz_1)\cpfp{n_2}{m_2}(z_2,\bz_2)
\cpfp{n_3}{m_3}(z_3,\bz_3)\Phi^{h_4}_{-q_4}(z_4,\bz_4)\rangle
\end{align*}
to find the structure constants of OP algebra via the corresponding 
three point correlation functions. The three point function of
interest is given by
\begin{align}
\langle
\cpfp{n_1}{m_1}(z_1,\bz_1)
\cpfp{n_2}{m_2}(z_2,\bz_2)
\Phi_{-q_3}^{h_3^+}(z_3,\bz_3)
\rangle=
C(n_1,m_1;n_2,m_2;n_3,m_3)\prod_{i<j}\vert
z_{ij}\vert^{-2\gamma_{ij}(h)}\quad,
\end{align}
where $q_3=q_1+q_2$ and the structure constants $C$ are written as
\begin{align*}
\begin{split}
C(n_1,m_1;n_2,m_2;n_3,m_3) &=C^+(n_1,m_1;n_2,m_2;n_3,m_3)
\frac{\prod_{i=1}^2 \gm [1+\prh(h_1+h_2+h_3-2h_i)]}
{\prod_{i=1}^3\gm [1+2\prh h_i]}\times\\
&\times\frac{\gm [1+\prh(h_1+h_2+h_3)]}
{\gm[1+\prh(h_3-h_1-h_2)]}\quad,\\
\end{split}
\end{align*}
with the coefficients $C^+$ defined in (2.13). The $\Gamma$-functions
come from the normalization factor $\mathcal N(j_3,-j_1-j_2)$ as well
as multiple integral over $x_i$. The latter is computed 
in Appendix C. Note that after setting $n_1=m_1=1$, the
integrals over $u_i\,,\,\,w_i$ are eliminated and the integral over
$x_1$ is decoupled, so the only multiple integral of interest is an
integral over $x_i\,,\,\,i=\{2,3,4\}$.

It is easy to see from (3.6) that the chiral primary fields
don't form a closed OP algebra at the rational level $k=-2+p/q$, with 
the coprime integers $p$ and $q$. The only exception is the unitary
series which correspond to $q=1$.

It should be noted that (3.6) represents the three point functions when all
conformal dimensions are parameterized by $h^+$. There
are, of course, three point functions with $h^-$. This is similar for the
case of the $SL(2)$ degenerate conformal fields theories (see ref.\cite{A1}).
The fusion rules then become
\begin{align*}
n_3=n_1+n_2-1\quad ,\quad m_3=m_1+m_2-1\quad
\end{align*}
and
\begin{align*} 
\begin{cases}
\vert n_{12}\vert +1\,\leq\, n_3\leq\text{min}\,(n_1+n_2-2,\,
2q-n_1-n_2+1)\,,\,\,\text{with}\,\,\,\Delta n_3=1\quad,\\
\vert m_{12}\vert +1\,\leq\, m_3\leq\text{min}\,(m_1+m_2-1,\,
2p-m_1-m_2-1)\,,\,\,\text{with}\,\,\,\Delta m_3=2\quad.
\end{cases}
\end{align*}
In above, only the first selection rule corresponds to the 
primary chiral field. As to the others, they correspond to the primary fields
which are no longer chiral. It is due to the U(1) conservation law 
$q_3=q_1+q_2$.
  
\vspace{.3cm}
{\bf 3.2 Chiral rings}

\vspace{.3cm}
The results of section 3.1 are forced me to look for new objects which
have a ring structure. In attempting to do this it is advantageous to
use operators introduced by Moore and Reshetikhin \cite{MR}\footnote{A
similar construction was also considered by Cremmer, Gervais and
Roussel \cite{G1}.}. 

According to \cite{MR} define holomorphic vertex operators, 
${}^{\alpha}\Phi^h_q(z)$, associated to a triple $(h,q,\alpha)$, where
$h$ and $q$ are the conformal dimension and U(1) charge,
respectively. As to $\alpha$, it means a pair of states in the highest weight 
representations of the quantum groups $(U_qosp(2/1),U_qsl(2))$. In
fact, I need a structure which manages the fusion of $(n,m)$, 
i.e. $(U_qosp(2/1),U_qsl(2))$ (see (2.15) and ref.\cite{MF}). 

The $N=2$ primary fields are given by
\begin{align}
\psfa(z,\bz)=\sum_{\alpha}\,{}^{\alpha}\psfa(z)\,^{\alpha}\psfa(\bz)\quad.
\end{align}

New features induced by the quantum groups are the corresponding Wigner
symbols in correlation functions and the Clebsch-Gordan coefficients
in the OP expansions. This implies, in particular, that 3-point
functions of operators ${}^{\alpha}\psfa(z,\bz)={}^{\alpha}\psfa(z)
{}^{\alpha}\psfa(\bz)$ are given by
\begin{align}
\langle{}^{\alpha_1}\Phi^{h_1}_{q_1}(z_1,\bz_1)\,
{}^{\alpha_2}\Phi^{h_2}_{q_2}(z_2,\bz_2)\,
{}^{\alpha_3}\Phi^{h_3}_{q_3}(z_3,\bz_3)\rangle=
J(\alpha_1,\alpha_2,\alpha_3)\,C(h_1,q_1;h_2,q_2;h_3,q_3)
\prod_{i<j}\vert z_{ij}\vert^{-2\gamma_{ij}(h)}\,\,,
\end{align}
where $J(\alpha_1,\alpha_2,\alpha_3)$ are the squares of the Wigner
symbols, $C(h_1,q_1;h_2,q_2;h_3,q_3)$ are structure constants of the
OP algebra of the primary fields.

If one denotes the vertices corresponding to the primary chiral fields
by ${}^{\alpha}\cpf{n}{m}$, then the 3-point functions of interest are
\begin{align}
\begin{split}
\langle{}^{\alpha_1}\cpfp{n_1}{m_1}(z_1,\bz_1)\,&
{}^{\alpha_2}\cpfp{n_2}{m_2}(z_2,\bz_2)\,
{}^{\alpha_3}\Phi^{h_3^{\pm}}_{-q_3}(z_3,\bz_3)\rangle=\\
&=J(\alpha_1,\alpha_2,\alpha_3)C(n_1,m_1;n_2,m_2;n_3,m_3)
\prod_{i<j}\vert z_{ij}\vert^{-2\gamma_{ij}(h)}\quad,
\end{split}
\end{align}
with 
\begin{align*}
\begin{split}
C(n_1,m_1;n_2,m_2;n_3,m_3) &=C^{\pm}(n_1,m_1;n_2,m_2;n_3,m_3)
\frac{\prod_{i=1}^2 \gm [1+\prh(h_1+h_2+h_3-2h_i)]}
{\prod_{i=1}^3\gm [1+2\prh h_i]}\times\\
&\times\frac{\gm [1+\prh(h_1+h_2+h_3)]}
{\gm[1+\prh(h_3-h_1-h_2)]}\quad,\\
\end{split}
\end{align*}
The coefficients $C^{\pm}$ depend on the parameterization of $h_3$
namely, the sign plus means
$h_3=h^+_{n_3.m_3}$, the sign minus - $h_3=h^-_{n_3.m_3}$ (see (2.13) 
and (2.14) for details).

If the states $\alpha_i\,,\,i=\{1,2\}\,$, are the highest weight
vectors then it is easy to find all non-zero correlation functions. In
the case of the rational level $k=-2+p/q$ 
the field ${}^{\alpha_3}\Phi^{h_3}_{-q_3}$ is uniquely
determined by
\begin{alignat*}{3}
n_3 &=n_1+n_2-1\quad &, \quad & m_3=m_1+m_2-1\quad &,\quad &1\leq n_3\leq
q\quad,\quad 1\leq m_3\leq p-1\quad, \\
h_3 &=h_1+h_2\quad &, \quad & q_3=q_1+q_2\quad &,\quad
&\alpha_3\,-\,\text{a pair of the lowest weight vectors}.
\end{alignat*}
The above result implies that the operators ${}^{\alpha}\cpf{n}{m}$
generate the ring
\begin{align}
{}^{\alpha_1}\Phi_{n_1.m_1}\times{}^{\alpha_2}\Phi_{n_2.m_2}=
\begin{cases}
{}^{\alpha_3}\Phi_{n_1+n_2-1.m_1+m_2-1}&\,,\quad n_1+n_2-1\leq
q\,,\quad m_1+m_2\leq p\,,\\
0&\,,\quad n_1+n_2-1>q\,,\quad m_1+m_2>p\,.
\end{cases}
\end{align}
At this point a few comments are in order:
\newline(i) Because the highest weights $(j,j^{\prime})$ of 
$(U_qosp(2/1),U_qsl(2))$ are
expressed in terms of $(n,m)$ as 
$(j=\frac{n-1}{2},j^{\prime}=\frac{m-1}{2})$ \cite{MF}, the Wigner
symbol (Clebsch-Gordan coefficient) provides $j_3=j_1+j_2\,,\,
j^{\prime}_3=j^{\prime}_1+j^{\prime}_2$ or
$n_3=n_1+n_2-1\,,\,m_3=m_1+m_2-1$ \cite{KR}. 
\newline (ii) One can use the relation between the chiral primary
fields and the highest weight vectors of $\2$ in order to see that
in a general case the chiral primary fields don't form the closed OP
algebra because the corresponding highest weight vectors don't do this
\cite{BFe}. However if one doesn't use screening operators,
that means that only the highest weight vectors of the quantum group
are allowed \cite{GS}, the fusion of $(n,m)$ is precisely 
$n_1\times n_2\rightarrow n_1+n_2-1\,,
\,m_1\times m_2\rightarrow m_1+m_2-1$.
\newline (iii) The operators ${}^{\alpha}\cpf{n}{m}$ which define the
ring obey the OP expansions (2.20) as well as (3.1), i.e. they are
primary and chiral.

\section{Conclusions and remarks}
\renewcommand{\theequation}{4.\arabic{equation}}
\setcounter{equation}{0}

First, let me say a few words about the results.

In this work I have found the relation between the $SL(2)$ degenerate
conformal field theories on one side and some $N=2$ discrete
superconformal series on the other side. This generalized fermionic
construction allows me to investigate the properties of the primary
chiral fields in the $N=2$ models. As a result, the OP algebra of such
fields was computed. It turned out that the primary chiral fields
don't generate the ring. The origin of the disaster is the
non-unitarity of the models. Next the Moore-Reshetikhin operators were
introduced to solve the problem. This solution gives a strong evidence
that a quantum group underlies the ring. It is disguised in the
unitary case in virtue of the U(1) conservation law, but it is becomes
clear in the non-unitary case. The experience with the fermionic
construction also shows that one has to take into account more exotic
modules over $\2$ to recover the all highest weight modules over $N=2$
(see point (iii) below for details). 

Let me conclude by mentioning some open problems:
\newline(i) It is clear that techniques developed in sections 2.1 and
2.2 allow one to consider any four point function
\begin{align}
\langle\,\prod_{i=1}^{4}\Phi^{h_i}_{q_i}(z_i,\bz_i)\,\rangle
=\prod_{i<j}^4\vert z_{ij}\vert^{\frac{-4\mu_i\mu_j}{k+2}}
\prod_{i=1}^4 \mathcal N^{-1}(j_i,\mu_i)
\oint_{C_i}\frac{dx_i}{x_i^{1+j_i-\mu_i}}
\oint_{\bar C_i}\frac{d\bx_i}{\bx_i^{1+j_i-\mu_i}}
\langle\,\prod_{i=1}^{4}\Phi^{j_i}(x_i,\bx_i,z_i,\bz_i)\,\rangle\quad,
\end{align}
with $h_i\,,\,\,q_i$ defined in (2.25).
\newline The correct contours $C_i(\bar C_i)$, for a particular
conformal block, should be chosen by the correct singularities at
$z_{ij}\rightarrow 0$, which should fit to an OP algebra obtained by
setting $n_1=m_1=1$. An exact prescription, for picking up the correct
contours is lacking at this time.

In fact, the problem is closely connected with generalized
Dotsenko-Fateev integrals. In the simplest case such integrals look
like
\begin{align}
I(a_1,a_2;b_1,b_2;c)=\prod_{i=1}^2\oint_{C_i}\,dx_i\,x_i^{a_i}
(1-x_i)^{b_i}\,x_{12}^c
\quad ,
\end{align}
with some real parameters $a_i\,,\,\,b_i\,,\,\,c$.
\newline Note that the integral (4.2) reduces to the Dotsenko-Fateev
one under $a_1=a_2$ and $b_1=b_2$.

I leave the analysis of these problems for future study.
\newline(ii) The second problem is interesting too. It concerns the
conjecture that $N=2$ superconformal field theories in two dimensions are
critical points of super-renormalizable Landau-Ginzburg (LG)
models. This conjecture followed a discussion of usual minimal models
(N=0) by Zamolodchikov \cite{Zam} and in the context of the $N=2$
minimal models was further developed by many authors (see
e.g. \cite{MVW} 
and refs therein). In the case of the $N=2$ discrete 
series it seems natural to follow the same procedure. Introducing two
chiral fields $X\,,\,\,Y$ which correspond to the fundamental fields
$\Phi_{1.2}$ and $\Phi_{2.1}$, one can write down an equation for the
superpotential $W(X,Y)$
\begin{align}
W(X,Y)=\frac{1}{k+2}X\frac{\partial}{\partial X}W(X,Y)-
Y\frac{\partial}{\partial Y}W(X,Y)\quad ,\quad k+2=\frac{p}{q}\quad .
\end{align}
However this equation has an infinite set of solutions. The solution
consistent with the $N=2$ minimal models $(q=1)$ can be written in the
form
\begin{align}
W(X,Y)=X^pY^{q-1}+\tilde W(X,Y)\quad ,\quad \tilde W(X,Y)\vert_{q=1}=0
\quad(\text{or}\,\,\tilde W(X,Y)\vert_{q=1}=f(Y))\quad. 
\end{align}
The main problem here is to find $\tilde W(X,Y)$. To do this one can
try to use the $\varepsilon$-expansion as it was done by Howe and West
in the case of the minimal models 
\cite{HW}. On the other hand, it
would be interesting to apply the Witten elliptic genus calculations
\cite{Wit} to the problem at hand. A natural question which also
arises: which algebraic varieties do give $W(X,Y)$? They are well-known 
for the minimal models (see e.g. \cite{S1}).
\newline(iii) One has seen in section 3.1 that it is not enough to
use only the highest(lowest) weight representations of $\2$ as
well as intermediate ones to describe all highest weight 
representations of $N=2$ for the discrete series. The first series
$h^I$ is recovered by considering modules over $\2$ with
$\mu=-j-1$. Such modules contain two parts: non-normalizable states
and normalizable ones. In the context of free field representations of
$\2$ similar modules were discussed in \cite{AF}. They reveal an
interesting submodule structure which is a mixture of the Verma and
Wakimoto structures. However a conformal field theory with primary
fields correspond to these modules is lacking at the moment. 

It should be noted that a similar problem is considered from the
mathematical point of view in \cite{AMS} where an equivalence between
some categories of modules over $\2$ and topological $N=2$ algebra is 
proven. The latter is closely connected with the standard
$N=2$ algebra and its chiral rings (see e.g. \cite{DVV} for details).
\newline(iv) An immediate consequence of section 3.1 is that a quantum
group structure underlying the $N=2$ discrete series is larger then
$(U_qosp(2/1),U_qsl(2))$. It is due to contributions from
$h^I_2\,,\,\,n=1$ and $h^{II}$ sectors. The problem is to find it exactly.

{\bf Acknowledgments.} I am indebted to J.-L.Gervais for useful
discussions. I am also grateful R.Schimmrigk and
A.Semikhatov for sending me their papers, and G.Lopes Cardoso for
reading the manuscript. The hospitality extended to
me at Laboratoire de Physique Th\'eorique de l'\'Ecole Normale Sup\'erieure,
where this work was done, is acknowledged. This research was supported
in part by Landau-ENS Jumelage, INTAS grant 94-4720, and by 
Russian Basic Research Foundation under grant 96-02-16507.

\vspace{.3cm}
\appendix{{\bf Appendix A}}
\renewcommand{\theequation}{A.\arabic{equation}}
\setcounter{equation}{0}

\vspace{.3cm}
The purpose of this appendix is to compute the normalization factors
relevant for the primary fields (2.16).

Let me normalize the two-point function as
\begin{align}
\langle\sn{\mu}{\mu}(z_1,\bz_1)
\Phi^j_{-\mu.-\bar \mu}(z_2,\bz_2)\rangle=
\frac{(-)^{\bar \mu -\mu}}{\vert z_{12}\vert ^{2\Delta_j}}\quad.
\end{align}

Using the corresponding normalization the $\pfc(x,\bx,z,\bz)$ fields 
(see (2.10)) one easily gets
\begin{align}
(-)^{\bar \mu-\mu}\mathcal N(j,\mu,\bar \mu)
\mathcal N(j,-\mu,-\bar \mu)=
\oint_{C_1}\oint_{C_2}dx_1dx_2\,
x_1^{\mu-1-j}x_2^{-\mu-1-j}x_{12}^{2j}\,\times\, (c.c)\quad. 
\end{align}
In above $(c.c)$ means integrals with $x_i\rightarrow\bar
x_i\,\,\text{and}
\,\,\mu\rightarrow\bar \mu\,$. 

The substitutions $x_2=tx_1\,\,\text{and}\,\, \bar x_2=\bar t\bar x_1$
lead to\footnote{I use the following normalization: $\oint_{C_0}\frac{dz}{z}=1.$}
\begin{align}
(-)^{\bar \mu-\mu}\mathcal N(j,\mu,\bar \mu)
\mathcal N(j,-\mu,-\bar \mu)=
\oint_{C_t}dt\,t^{-\mu-1-j}(1-t)^{2j}\,\times\,(c.c)\quad.
\end{align}
\unitlength=1mm
\special{em:linewidth 0.4pt}
\linethickness{0.4pt}
\begin{picture}(95.33,45.00)
\put(101.00,25.00){\circle*{1.33}}
\put(75.00,25.00){\circle*{1.33}}
\bezier{184}(101.00,25.00)(70.67,38.00)(70.00,25.00)
\bezier{180}(101.00,25.00)(72.00,12.00)(70.00,25.00)
\bezier{256}(101.00,25.00)(62.33,45.00)(61.00,25.00)
\bezier{256}(101.00,25.00)(62.33,5.00)(61.00,25.00)
\put(66.00,34.00){\vector(-4,-3){2.00}}
\put(77.33,31.50){\vector(1,0){2.67}}
%\put(62.00,25.00){\circle*{0.67}}
%\put(64.67,25.00){\circle*{0.67}}
%\put(67.33,25.00){\circle*{0.67}}
\put(105.33,25.00){\makebox(0,0)[cc]{1}}
\put(80.00,25.00){\makebox(0,0)[cc]{0}}
\put(59.67,32.67){\makebox(0,0)[cc]{$t$}}
\put(69.00,31.00){\makebox(0,0)[cc]{$\bar t$}}
\end{picture}
\vspace{-1cm}
\begin{center}
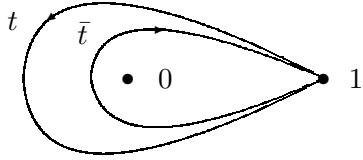
Fig.1 Contours used in the definition of the normalization factors
$\mathcal N(j,\mu,\bar \mu)$.
\end{center}

\vspace{.3cm}
Choosing the contours $C_t\,\,\text{and}\,\,\bar C_{\bar t}$ as shown
in Fig.1 and using the definition of the B-function, one finds
\begin{align}
\mathcal N(j,\mu,\bar \mu)\mathcal N(j,-\mu,-\bar \mu)=
\Gamma ^2[2j+1]\{\gm[1+j+\mu]\gm[1+j-\mu]
\gm[1+j+\bar \mu]\gm[1+j-\bar \mu]\}^{-1} \quad.
\end{align}

Finally, the normalization factors are given by
\begin{align}
\mathcal N(j,\mu,\bar \mu)=\mathcal N(j,-\mu,-\bar \mu)=
\gm[2j+1]\{\gm[1+j+\mu]\gm[1+j-\mu]
\gm[1+j+\bar \mu]\gm[1+j-\bar \mu]\}^{-\frac{1}{2}} \quad.
\end{align}

For ''A'' series (A.5) reduces to
\begin{align}
\mathcal N(j,\mu)=\frac{\gm[2j+1]}{\gm[1+j+\mu]\gm[1+j-\mu]} \quad.
\end{align}

\vspace{.3cm}
\appendix{{\bf Appendix B}
\renewcommand{\theequation}{B.\arabic{equation}}
\setcounter{equation}{0}

\vspace{.3cm}
It turns out that it is easy to show that the primary fields of the
$N=2$ models parameterized by the first series $h^I$ form a closed OP
algebra, i.e. there is decoupling of the second series
$h^{II}$. To do this, I will use the free field representation of
$N=2$ algebra constructed by Yu and Zheng \cite{YZ}.

In the NS sector the holomorphic part is described by two free scalar
chiral superfields $\boldsymbol{\phi}^{\pm}$ coupled to a background
charge. In terms of the component fields, complex scalars and
fermions, one has
\begin{align*}
\sp^+(\boz)=\varphi^+(z)+\sqrt{2}\theta^-\psi^+(z)
+\theta^-\theta^+\partial\varphi^+(z)\quad ,\quad
\sp^-(\boz)=\varphi^-(z)+\sqrt{2}\theta^+\psi^-(z)
-\theta^-\theta^+\partial\varphi^-(z)\,\,,
\end{align*}
where $\boz=(z,\theta^+,\theta^-)$ is a point on the supersphere.
\newline
The chirality means $D^-\sp^+(\boz)=D^+\sp^-(\boz)=0$ with the
superderivatives
$D^{\pm}=\frac{\partial}{\partial\theta^{\mp}}+\theta^{\pm}
\frac{\partial}{\partial z}$. The two point functions of the component
fields are normalized as 
\begin{align*}
\langle\,\varphi^+(z_1)\varphi^-(z_2)\,\rangle=-\log z_{12}\quad ,\quad
\langle\,\psi^+(z_1)\psi^-(z_2)\,\rangle=\frac{1}{z_{12}}\quad.
\end{align*}

The holomorphic supercurrent is given by
\begin{align}
\mathcal J(\boz)=\frac{1}{2}D^+\sp^+D^-\sp^-(\boz)
+\frac{i}{2}\boa_0\bigl(\partial\sp^+(\boz)-
\partial\sp^-(\boz)\bigr)\quad.
\end{align}
It is an exercise on OP expansions to check that $\mathcal J$ has the
following OP algebra
\begin{align}
\mathcal J(\boz_1)\mathcal J(\boz_2)=
\frac{c_2/4}{\boz_{12}^2}+
\biggl(\frac{\theta^-_{12}\theta^+_{12}}{\boz_{12}^2}+
\frac{1}{2}\frac{\theta^-_{12}}{\boz_{12}}D^+ -
\frac{1}{2}\frac{\theta^+_{12}}{\boz_{12}}D^- +
\frac{\theta^-_{12}\theta^+_{12}}{\boz_{12}}\frac{\partial}{\partial
z_2}\biggr)\mathcal J(\boz_2)+\dots\quad,
\end{align}
with
$\boz_{12}=z_{12}-\theta^-_1\theta^+_2-\theta^+_1\theta^-_2\,\,\text{and}
\,\,c_2=1-2\boa_0^2$.

The primary fields are given by
\begin{align}
V_{\boa^+.\boa^-}(\boz)=e^{i\bigl(\boa^+\sp^-(\boz)+
\boa^-\sp^+(\boz)\bigr)}\quad.
\end{align}
They have the following OP expansions with the supercurrent
\begin{align}
\mathcal J(\boz_1)V_{\boa^+.\boa^-}(\boz_2)=
\biggl(h\frac{\theta^-_{12}\theta^+_{12}}{\boz_{12}^2}+
\frac{1}{2}\frac{\theta^-_{12}}{\boz_{12}}D^+ -
\frac{1}{2}\frac{\theta^+_{12}}{\boz_{12}}D^- +
\frac{\theta^-_{12}\theta^+_{12}}{\boz_{12}}\frac{\partial}{\partial z_2}+
\frac{q}{\boz_{12}}\biggr) V_{\boa^+.\boa^-}(\boz_2)+\dots\,\,,
\end{align}
where the conformal dimension $h=\boa^+\boa^--\frac{\boa_0}{2}(\boa^++
\boa^-)\,\,\text{and U(1) charge}\,\,q=\frac{\boa_0}{2}(\boa^+-\boa^-)$.

The screening operators are expressed as
\begin{align}
S=\oint dzd\theta^+d\theta^-V_{\ap.\ap}(\boz)\quad,\quad
F^+=\oint dzd\theta^+V_{\am.0}(\boz)\quad ,\quad
F^-=\oint dzd\theta^-V_{0.\am}(\boz)\quad ,\quad
\end{align}
with $\am=-\sqrt{k+2}\,,\,\,\ap\am=-1$.

In the free field representation the first series $h^I$ is described
by \cite{YZ} 
\begin{align}
\boa^++\boa^-+n\am+m\ap=0\quad,\quad
q=\frac{\boa_0}{2}(\boa^+-\boa^-)\quad.
\end{align}
As to the second, it corresponds to
\begin{align}
\boa^+=-\frac{1}{2}m\ap\quad,\quad
q=\frac{\boa_0}{2}(\boa^+-\boa^-)\quad.
\end{align}

Now let me look at the three point function which contains two
primaries from the first series $h^I$ and one from the second series 
$h^{II}$. The free field representation results in
\begin{align}
\langle \Phi_{q_1}^{h^I_1}(\boz_1)\Phi_{q_2}^{h^I_2}(\boz_2)
\Phi_{q_3}^{h^{II}_3}(\boz_3)\rangle=
\langle V_{\boa_1^+.\boa_1^-}(\boz_1)
\Tilde V_{\boa_2^+.\boa_2^-}(\boz_2)
V_{\boa_3^+.\boa_3^-}(\boz_3)
\prod_{i=1}^s S_i\prod_{j=1}^{f^+} F_j^+\prod_{l=1}^{f^-} F_l^-\rangle\,\,,
\end{align}
with a conjugate vertex operator 
$\Tilde V_{\boa^+.\boa^-}(\boz)=
V_{\boa_0-\boa^+.\boa_0-\boa^-}(\boz)$ \cite{DF,YZ} and
$\{s,f^{\pm}\}\in\N.$.

The balance of charges (zero modes) leads to
\begin{align} 
\begin{cases}
\boa_1^+ +\boa_1^- -\boa_2^+ -\boa_2^- +\boa_3^+ +\boa_3^-
+2s\ap +(f^+ +f^-)\am=0\quad,\\
-\boa_1^+ +\boa_1^- +\boa_2^+ -\boa_2^- -\boa_3^+ +\boa_3^-
+(f^+ -f^-)\am=0\quad.
\end{cases}
\end{align}
Taking the first equation, combined with (B.6) and (B.7), I find
\begin{align}
q_3=\frac{1}{2}(n_1-n_2-f^+-f^-)+\frac{1}{2}(-m_1+m_2-m_3+2s)(k+2)^{-1}\quad,
\end{align}
which implies that the conformal block isn't zero if the charge $q_3$
is quantized like the weights (2.7) up to a factor $k+2$. Since in this
case the series
$h^{II}$ is equivalent to 
the $h^I$ one namely, $h_p^{II}=h^I_{\alpha.p+\beta}
\,\,\text{with}\,\,q=\frac{1-\alpha}{2}+\frac{\beta}{2}(k+2)^{-1}
\,;\,\,\{\alpha,\beta\}\in\N$, it means decoupling of the second series.

In the above, I have considered the 3-point conformal block. However, the
generalization to a n-point one is straightforward.

\vspace{.3cm}
\appendix{{\bf Appendix C}
\renewcommand{\theequation}{C.\arabic{equation}}
\setcounter{equation}{0}

\vspace{.3cm}
In this Appendix I will compute a multiple contour integral used in
sections 3.1, 3.2 in order to build the ring structure.

Let me consider an integral
\begin{align}
I(a,b,c)=\prod_{i=1}^2\oint_{C_{x_i}}\,\frac{dx_i}{x_i}
\oint_{C_{x_3}}\,\frac{dx_3}{x_3^{1+a+b+c}}\,x_{31}^ax_{32}^bx_{12}^c
\,\,\times\,\,(c.c)\quad ,
\end{align}
with some real parameters $a\,,\,\,b\,,\,\,c$.

The substitutions $x_1=t_1x_3\,,\,\,x_2=t_2x_3\,(\bx_1=\bar
t_1\bx_3\,,\,\,\bx_2=\bar t_2\bx_3)$ lead to
\begin{align}
I(a,b,c)=\prod_{i=1}^2\oint_{C_i}\,\frac{dt_i}{t_i}
\,(1-t_1)^a(1-t_2)^bt_{12}^c\,\,\times\,\,(c.c)
\quad .
\end{align}
Note that the holomorphic integral depends on contours $C_i$. In a
general case $a\not = b$ there is no symmetry $C_1\rightarrow C_2\,,\,\,
C_2\rightarrow C_1$, i.e. the integral is not the Dotsenko-Fateev
type. At the case at hand one has $C_1\rightarrow C_2\,,
\,\,C_2\rightarrow C_1\,,\,\,a\rightarrow b\,,\,\,b\rightarrow a$. Due
to this reason there are two possibilities for contours $C_i$ 
namely\footnote{As in the case of Appendix A contours $\bar C_i$ are
taken in such way to cancel relative phases of integrals over $x_i$
and $\bx_i$.},

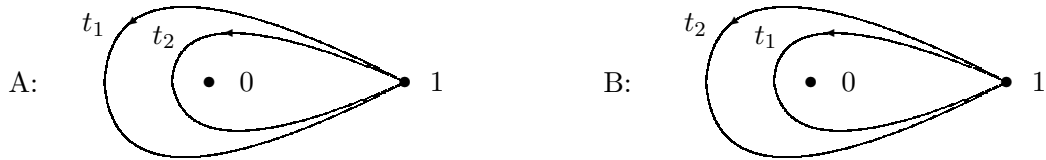
\unitlength=1mm
\special{em:linewidth 0.4pt}
\linethickness{0.4pt}
\begin{picture}(50.33,45.00)
\put(56.00,25.00){\circle*{1.33}}
\put(30.00,25.00){\circle*{1.33}}
\bezier{184}(56.00,25.00)(25.67,38.00)(25.00,25.00)
\bezier{180}(56.00,25.00)(27.00,12.00)(25.00,25.00)
\bezier{256}(56.00,25.00)(17.33,45.00)(16.00,25.00)
\bezier{256}(56.00,25.00)(17.33,5.00)(16.00,25.00)
\put(21.00,34.00){\vector(-4,-3){2.00}}
\put(34.33,31.50){\vector(-1,0){2.67}}
%\put(62.00,25.00){\circle*{0.67}}
%\put(64.67,25.00){\circle*{0.67}}
%\put(67.33,25.00){\circle*{0.67}}
\put(60.33,25.00){\makebox(0,0)[cc]{1}}
\put(35.00,25.00){\makebox(0,0)[cc]{0}}
\put(14.67,32.67){\makebox(0,0)[cc]{$t_1$}}
\put(24.00,31.00){\makebox(0,0)[cc]{$t_2$}}
\put(136.00,25.00){\circle*{1.33}}
\put(110.00,25.00){\circle*{1.33}}
\bezier{184}(136.00,25.00)(105.67,38.00)(105.00,25.00)
\bezier{180}(136.00,25.00)(107.00,12.00)(105.00,25.00)
\bezier{256}(136.00,25.00)(97.33,45.00)(96.00,25.00)
\bezier{256}(136.00,25.00)(97.33,5.00)(96.00,25.00)
\put(101.00,34.00){\vector(-4,-3){2.00}}
\put(114.33,31.50){\vector(-1,0){2.67}}
%\put(62.00,25.00){\circle*{0.67}}
%\put(64.67,25.00){\circle*{0.67}}
%\put(67.33,25.00){\circle*{0.67}}
\put(140.33,25.00){\makebox(0,0)[cc]{1}}
\put(115.00,25.00){\makebox(0,0)[cc]{0}}
\put(94.67,32.67){\makebox(0,0)[cc]{$t_2$}}
\put(104.00,31.00){\makebox(0,0)[cc]{$t_1$}}
\put(5.33,25.00){\makebox(0,0)[cc]{A:}}
\put(84.33,25.00){\makebox(0,0)[cc]{B:}}
\end{picture}
\vspace{-1cm}
\begin{center}
Fig.2 Basic contours $C_i$ for the integral (C.2).
\end{center}

\vspace{.3cm}
As a result, one has
\begin{align}
I^{(A)}(a,b,c)=\frac{\gm^2[1+a]}{\gm^2[1-c]\gm^2[1+a+c]}\quad ,\quad
I^{(B)}(a,b,c)=\frac{\gm^2[1+b]}{\gm^2[1-c]\gm^2[1+b+c]}\quad .
\end{align}

Since all three point functions considered in sections 3.1 and 3.2 are
symmetric under $(n_1,m_1)\rightarrow (n_2,m_2)\,,
\,\,(n_2,m_2)\rightarrow (n_1,m_1)$ one is free to symmetrize a factor
which comes from $\mathcal N(j_3,-j_1-j_2)$ as well as $I^{(i)}$. I
use the following ansatz
\begin{align}
\mathcal N^{-1}(j_3,-j_1-j_2)I^{(i)}\rightarrow
\frac{\gm[1+j_1+j_2+j_3]}{\gm[1-j_1-j_2+j_3]}
\frac{\prod_{i=1}^2\gm[1+j_1+j_2+j_3-2j_i]}
{\prod_{i=1}^3\gm[1+2j_i]}\quad.
\end{align}
It should be stressed that the first factor in the above is universal
under any symmetrization prescription due to its explicit symmetry. On the
other hand, it is the most important one since this is an origin for a
truncation of fusion rules. From this point of view the results
(fusion rules) are independent on the contours $C_i(\bar C_i)$.

\end{document}